\documentstyle[sprocl]{article}
\input{psfig}
\def\Journal#1#2#3#4{{#1} {\bf #2}, #3 (#4)}

\def\PLB{{\em Phys. Lett.}  B}
\def\PRL{\em Phys. Rev. Lett.}
\def\PRD{{\em Phys. Rev.} D}

\def\be{\begin{equation}}
\def\ee{\end{equation}}
\def\bea{\begin{eqnarray}}
\def\eea{\end{eqnarray}}

\begin{document}

\title{Status and Future Prospects in Searches for New Interactions in Neutron and
Nuclear Beta-decay, Muon- and Pion-decay}

\author{ J. DEUTSCH\footnote{The numerous colleagues who kindly helped providing
information and criticism are thankfully acknowledged at the end of the text} }

\address{Universit\'e catholique de Louvain, Institut de Physique,\\
2 chemin du Cyclotron, B-1348 Louvain-la-Neuve, Belgium\\ e-mail : deutsch@fynu.ucl.ac.be}


\maketitle\abstracts{
The interest, the status and the perspectives of various experiments in neutron
and nuclear beta-decay, muon-decay and pion-decays are discussed.  The talk is segmented
into a discussion of the decay-rates and of the energy-spectra and correlations.
The impact on various scenarios of "new physics" is briefly mentioned~; left-right
symmetric models are discussed in more detail and the informations gained from the
considered experiments is compared to those from other sources.}

\section{Decay-rates}
\subsection{Muon-decay}\label{subsec:prod}

\noindent The muon decay rate (1/$\tau_{\mu}$) is directly
related to the Fermi coupling constant $G_F$, a fundamental
quantity of the Standard Model : (1/$\tau_{\mu}$) = $G_{F}$K where
the factor K contains mass-terms and radiative corrections \cite{Sirlin80}. It was
stressed by L.B. Okun that $G_F$ is an important quantity as it is
directly related to the vacuum expectation value of the Higgs
field which sets the fundamental mass-scale of the Standard Model \cite{Okun80}.
The relative precision to which $G_F$ was known was, up to
recently, 17 ppm, where 9 ppm is of experimental origin and 15 ppm
was an estimate of the 2-loop corrections in the factor K.
Recently, these corrections were evaluated to high precision \cite{Ritbergen98} and
by now the imprecision in $G_F$ is dominated by the experimental
error.

\noindent Various groups plan to remesure the muon decay constant
with improved precision.  In particular, a
Bologna-CERN-ETH-PSI-ULB/LBL collaboration developed a
fine-grained fiber scintillation target and will use the unique
muon beam of PSI to measure the decay constant to a precision of 2
ps, an order of magnitude improvement over the present one \cite{Kirkby98}.

\noindent It should be noted that the relation between the decay
constant and $G_F$ could be modified by deviations from the
Standard Model \cite{Fetscher95}. E.g. a right-handed scalar coupling of a relative
importance of p would modify $G_F$ by 0.02 p !  As the transverse
polarization of the decay positrons is sensitive to such
deviations from the Standard Model, an experiment is performed
presently at PSI to measure p or to set an upper limit to it at
the $\%$-level \cite{Fetscher94}.

\subsection{Pion beta-decay}

\noindent The importance of the precision study of superallowed
pure Fermi decays to check the unitarity of the
Cabibbo-Kobayashi-Maskawa matrix was stressed by I. Towner in this
conference \cite{Towner}. He noted also that the investigated nuclear
beta-decays results in a 2.2 standard deviation from unitarity. He
stressed also that the error of this determination is dominated by
radiative and nuclear structure corrections.

\noindent In order to overcome the need for these corrections, an
experiment is under way to determine the rate of the superallowed
Fermi-decay of the pion \cite{Pocanic89}. A measurement of this branching ratio,
with a precision comparable to the ones obtained in nuclear
beta-decays, is very difficult as this decay branch is only
$10^{-8}$. Having constructed a spectrometer of nearly 4 $\pi$
coverage and profiting from the high pion flux at PSI, D. Pocanic
and his colleagues aim, in the first phase of the experiment, to
obtain a 0.5 $\%$ precision on the branching ratio.

\subsection{Ratio of the electron/muon decay-rate of the pion}

\noindent In the Standard V-A Model this ratio is
helicity-suppressed to the 10$^{-4}$ level and its precise
determination verifies not only the universality of the
electron/muon coupling strenghts, but also the absence of
particles beyond the Standard Model \cite{Britton94} \cite{Herczeg95}.

\noindent The combination of the TRIUMF \cite{Britton94} and PSI
\cite{Czapek92} precision experiments results in a ratio of
(1.2312 +/- 0.0037)10$^{-4}$ to be compared to the Standard Model
prediction of (1.2352 +/- 0.0005)10$^{-4}$. The comparison verifies the
universality of the electron-muon coupling (g$_e$/g$_{\mu}$ =
0.9989 +/- 0.0016) and excludes new particles with mass-bounds
in the TeV-region (cfr references quoted above). As the precision
of the test is now limited by the experimental contribution, a
6-fold improvement in the measurement is planed at TRIUMF
\cite{Numao}.

\subsection{Neutron life-time}

\noindent The decay rate of the neutron is an interesting quantity
because of its cosmological impact on the synthesis of the light
elements \cite{Byrne82} and because, combined with the electron
asymmetry, it provides, in the Standard Model, a direct
determination of the semileptonic vector- and axial-vector
coupling constants (cfr 2.3.1). The confrontation of the vector
coupling obtained by this procedure with the one extracted from
the decay rate of the superallowed Fermi transitions can observe
(or place constraints) on physics beyond the Standard Model (cfr
3.3).

\noindent P. Geltenbort communicated to this conference the most
recent result of L.N. Bondarenko et al. \cite{Geltenbort} :
$\tau_n$ = 885.4 +/- 0.9 (stat) +/- 0.4 (syst.) sec resulting in a
world average of $\tau_n$ = 885.8 +/- 0.9 sec. Improvement of this
precision are planed both at ILL \cite{Geltenbort_1} and at NIST
where a factor of 10 improvement is expected using superfluid
helium both to produce ultracold neutrons and to detect their
decay \cite{Doyle}.

\section{Spectra and Correlations}

\subsection{The Michel parameters in muon-decay}

The Michel-parameters are phenomenological quantities which
describe the various observables in muon-decay \cite{Fetscher95}.
They can be related to the leptonic coupling constants, only one
of them ($g^V_{LL}$) being non-zero in the Standard Model.  As a
consequence, precision measurements of these parameters test
physics beyond the Standard Model \cite{Herczeg86}, cfr. also section 3.3.

\noindent In addition to a measurement of the $\rho$ parameter by
the MEGA collaboration which is under analysis \cite{Mischke},
three different precision experiments are progressing (D. Gill et
al., TRIUMF Exp. E614, W. Fetscher et al., PSI Exp. R-94.10 and R.
Prieels et al., PSI Exp. R-97.06).  In table 1, their
precision aims are compared to the present precision of the Michel
parameters.

\begin{table}[h]
\caption{Present precision (in 10$^{-3}$) of the various Michel parameters
and precision aims of the ongoing experiments \label{tab:pres}}
\vspace{0.4cm}
\begin{center}
\begin{tabular}{|c|c|c|c|c|}
\hline
& & & & \\
& Part. Data 98 & TRIUMF E614 & PSI R-94.10 & PSI R-97.06 \\
\hline
$\rho$ & 3 & 0.1 & & \\
$P_{\mu} \xi$ & 9 & 0.14 & & \\
$\delta$ & 4 & 0.3 & & \\
$\xi \delta / \rho$ & 3 & & & \\
$\eta$ & 13 & 3 & $\sim$ 3 & \\
& & & & \\
P$_T$ (T-odd) & 23 &  & $\sim$ 5 & \\
P$_T$ (T-even) & 85 & & $\sim$ 30 & \\
P$_L$ : [$\xi$']  & 45 & & & \\
P$_L$ : [$\xi$''/($\xi$' $\xi$) ]& 360 & & & $\sim$ 2 \\
\hline
\end{tabular}
\end{center}
\end{table}

\subsection{Electron-neutrino directional correlations in beta-decay}

\noindent We refer to the paper of J.D. Jackson et al.
\cite{Jackson57} for the dependency of the various beta-decay
observables on the four-fermion coupling constants $C_i$ and
$C'_i$. In the Standard Model only $C_V$ = $C'_V$ = 1 and $C_A$ = $C'_A$
are non-zero and so precision measurements  of the various
observables can observe or constrain physics beyond the Standard
Model \cite{Herczeg95_1} \cite{Deutsch95}.

\noindent In particular, electron-neutrino directional correlation
measurements in pure Gamow-Teller (Fermi) transitions can observe
or constrain the absolute value of $C_{T(s)}$. Such couplings could
originate in the exchange of leptoquarks or charged scalar bosons
\cite{Herczeg95_1}. The pure Gamow-Teller decay of $^6$He was
investigated by C.H. Johnson {\it et al.} \cite{Johnson63}. The radiative
corrections to this decay were revisited recently \cite{Glueck98}
and it is now the corrected experimental value of the correlation
coefficient a = 0.3310 +/- 0.0030 which should be compared to the
one expected for a pure axial interaction : a = 0.3333.  The
experiment allows a tensor coupling-strength which can be up to
0.8 $\%$ (at 68 $\%$ CL) of the axial one.

\noindent In pure Fermi decays the correlation coefficient was
deduced from the recoil energy of the residual nucleus observing
either its gamma-decay \cite{Egorov97} or its proton-decay
\cite{Adelberger93}.  This latter technique was further improved
\cite{ISOLDE} and reached by now a statistical precision of 0.005
on the correlation coefficient.  The systematic contribution to
the error is still under investigation and variants of the
experiment are considered at other laboratories.

\noindent Because of the small recoil-energy to be observed,
optical or electromagnetic traps are particularly promising for
these experiments.  The most advanced project is the trapping of
$^{38m}$K at TRIUMF which will allow to constrain possible scalar
couplings \cite{Behr}.  Another similar experiment is in
preparation at LANL on $^{82}$Rb \cite{Freedman}.  Electromagnetic traps with
similar objectives are under construction at ANL \cite{Savard} and
ISOLDE \cite{Severijns}.  If the trapped nuclei are polarized,
they can also be used for asymmetry-measurements \cite{Freedman}.

\subsection{Polarization observables}

\subsubsection{Neutron beta-decay asymmetries}

\noindent Polarized neutrons allow to observe both the electron
and neutrino emission asymmetries ($A_n$ and $B_n$).  In the
Standard Model these asymmetries depend only from the ratio of the
axial- and vector coupling constants $\lambda$ = $G_A$/$G_V$, the
dependency of $A_n$ being stronger than that of $B_n$ :
$\sigma_A$/$\sigma_{\lambda}$ = 0.36 and $\sigma_B$/$\sigma_{\lambda}$
= 0.08.  As a consequence a measurement of $A_n$, combined with
that of the neutron life-time, allows a determination of $G_V$,
i.e. a verification of the unitarity of the
Cabibbo-Kobayashi-Maskawa matrix independently of nuclear
structure corrections (cfr 1.2). Using $G_V$ obtained from nuclear
beta-decay, it allows a control of left-right symmetric extensions
of the Standard Model (cfr 3.3). A measurement of $B_n$,
practically independent of $G_A$/$G_V$, is however better suited
for this second purpose.

\noindent Unfortunately the various results of the
$A_n$-measurements scatter beyond statistics.  Let us mention for
illustration the four latest (and most precise) results~: $A_n$ =
- 0.1160 +/- 0.0014 \cite{Schreckenbach}, $A_n$ = - 0.1189 +/-
0.0012 \cite{Liaux97}, $A_n$ = - 0.1135 +/- 0.0014
\cite{Kuznetsov98} and finally, presented at this meeting, $A_n$ =
- 0.1187 +/- 0.0008 \cite{Abele}. The weighted average of all the
available results is - 0.11721 +/- 0.00053 ; the $\chi^2$/n is
however too high : 3.3. A value with inflated errors (0.001) will
be used in section 3.3 for further analysis ; the situation is
however unsatisfactory. One of the critical points of the
experiment is the determination of the neutron polarization ;
experiments are under way at ILL to compare various methods to
determine this polarization \cite{Zimmer}. Also relative
precisions of about 1~$\%$ for $A_n$ are claimed but the
corrections to the measured values were of the order of 10~$\%$
(about 1$\%$ in the case of PERKEO II 98).  A novel and ambitious
experiment is started at LANSCE which, in its first phase, will
measure $A_n$ to 0.2 $\%$ with corrections of only 0.16 $\%$
\cite{Bowles}. The increase of neutron-flux expected from the
superfluid helium moderator (cfr 1.4) will further increase the
possibilities of the experiment and allow to measure also other
neutron-decay observables.

\noindent For $B_n$ a final result was  submitted
to this conference by I. Kuznetsov
{\it et al.}  ($B_n~=~0.9796~+/-~0.0044$) which results in a world
average of $B_n~=~0.9816~+/-~0.0040$.  This value is somewhat
smaller than the (V-A)-prediction ($\Delta~=~0.0064~+/-~0.0041$)
which has an impact on the discussions on right-handed currents
(section 3.3).

\noindent We shall not expand here upon the measurements of the
emission-asymmetry (and absolute
electron-polarization) measurements in nuclear decays
\cite{Deutsch95}. As these are absolute measurements, it is
difficult to obtain a good accuracy. Various corresponding
experiments are however in preparation (cfr our remarks on traps
in section 2.2).

\subsubsection{Beta-decay asymmetry/polarization correlations}

2.3.2.1. Nuclear beta-decay

\noindent A novel type of experiments was proposed by P. Quin and
T. Girard \cite{Quin89} which consists in comparing the
longitudinal polarization of the decay-electrons (positrons)
emitted in opposite directions by polarized nuclei. (A variant was
proposed by J. Govaerts et al. \cite{Govaerts95}, consisting in
the comparison of the electron-polarization for polarized and
unpolarized nuclei). These experiments are relative experiments
and so avoid the need to determine the absolute polarization of
the sample. They are, moreover, very sensitive to physics beyond
the Standard Model \cite{Govaerts95}. E.g. for a pure Gamow-Teller
transition and neglecting recoil order terms, the Standard Model
value of the polarization ratio $R_{0}$ takes the
form~\cite{Govaerts95}~\cite{Thomas97} :

\begin{equation}
R_0 = \frac{P_L(-J)}{P_L(J)} = \frac{1}{1 - A_{exp}}
\frac{\beta^2 (2-A_{exp}) - A_{exp}}{\beta^2 (2-A_{exp}) +
A_{exp}}
\end{equation}
where $\beta$ is the positron velocity and $A_{exp}$ is the
experimental asymmetry which, as we shall see, has not to be known
with good precision.  The ratio of the experimental value of R
compared to the Standard Model value $R_{0}$ is then :

\begin{equation}
\frac{R}{R_{0}} = 1 - k_{A} \frac{\Delta_{A}}
{1 + 4
\frac{A_{exp}}
{\beta^{2} (2 - A_{exp}) + A_{exp}} \Delta_{A}}
\end{equation}

where $\Delta_{A}$ is a measure of a deviation from the Standard
Model and

\begin{equation}
k_{A} = 8 \frac{\beta^2 A_{exp} (2 - A_{exp})}{\beta^4 (2 -
A_{exp})^2 - A^2_{exp}}
\end{equation}

is an enhancement factor which can be large indeed if $A_{exp}$ is
close to unity (i.e. if the transition is well chosen and the
polarization is sizeable). An enhancement factor of $k_A$ = 5-7 was
readily achieved in the study of $^{117}In$ \cite{Severijns93} \cite{Camps97}and
higher enhancement factors are hoped for in a similar experiment
in preparation at CERN-ISOLDE (CERN/ISC 96-11, ISC/P80).  Similar
precisions on $\Delta_A$(though with a smaller enhancement factor
$k_A$) were obtained at PSI with $^{12}N$ \cite{Allet96} \cite{Thomas97}. At the
level of precision obtained the impact of the nuclear structure
dependent recoil-order corrections is negligible.  The impact of
these measurements on left-right symmetric models will be
illustrated in section 3.3.

\noindent 2.3.2.2. Muon-decay

\noindent A similar situation can be expected in muon-decay with
the additional (experimental) bonus that muons are produced
strongly polarized in pion-decay at rest.  Indeed, for "backward"
positrons near the spectrum end-point (the reduced positron-energy
x=1), we have \cite{Govaerts92} :

\begin{equation}
P_L (x=1, cos \theta = -1) = \xi' + \frac{- P_{\mu} \xi}{1 - P_{\mu}}
\frac{\xi'' - \xi \xi'}{\xi}
\end{equation}

where the symbols $\xi$, $\xi'$ and $\xi''$ are Michel parameters
introduced already in section 2.1.  Note that the combination
($\xi'' - \xi \xi'/\xi$) is zero in the Standard Model and that the
enhancement factor  ($-P_{\mu} \xi$)/(1 - $P_{\mu}\xi$) can be
large for strongly polarized muons.  We mentioned already in
section 2.1 the PSI experiment R-95-06 of R. Prieels et al., which
will exploit this sensitivity to search physics beyond the
Standard Model.  This will be further illustrated in section 3.3.

\section{Impact on New Physics beyond the Standard Model}

\subsection{Exotic fermions, charged Higgs, s-lepton exchange}

\noindent We did not expand on these scenarios of physics beyond
the Standard Model and refer the reader to the recent review paper
of P. Herczeg \cite{Herczeg95_1} and to the many references provided
in it.

\subsection{Leptoquark exchange}

\noindent In view of the interest raised recently in scenarios
involving leptoquarks, let us stress (in addition of refering agin
to the review-paper of P. Herczeg), that the helicity-dependent
observables such as the ones discussed in section 2.3.2.1, allow
also to constrain masses and coupling-strength of various classes
of lepto-quarks \cite{Deutsch95} \cite{Quin96} \cite{Govaerts97}

\subsection{Left-right symmetric model}

\noindent Soon after the discovery of parity-violation in weak
interactions and its incorporation into the SU(2)$_L$x U(1)
gauge-group of the Standard Model, scenarios were proposed to
recover parity-symmetry at higher energies (for a discussion of
these attempts and corresponding references we refer to the
review-papers of P. Herczeg as well as J. Deutsch and P. Quin
already quoted).  For illustration we shall first consider the
so-called manifest left-right symmetric models, which contain only
two parameters and discuss in a second section the generalization
of these models.

\subsubsection{Manifest left-right symmetric models}

3.3.1.1. Nuclear beta-decay

\noindent In this model (refs. cfr. above) a second gauge boson is
introduced with a mass ($m_2$) larger than that of the observed W
($m_1$ = 80 GeV). This second gauge boson has a predominantly
right-handed coupling.  Formally the flavor coupling states $W_{L(R)}$
are written as

\begin{equation}
W_L = cos \zeta W_1 + sin \zeta W_2 \hspace*{1cm}
W_R = - sin \zeta W_1 + cos \zeta W_2
\end{equation}

(the mixing-angle $\zeta$ being small).  We shall introduce the
notation $\delta~=~(m_1/m_2)^2$.

Writing the left(right) handed fermion covariants as L(R), for
small mixing and $q^2 << m_1$ , we obtain a hamiltonian of the
following form :

\begin{eqnarray}
H       & =       & ( g^{2} / 8 )  \left\{
         \left[ \frac{cos^{2} \zeta }{ q^{2} + m^{2}_{1}}  + \frac{sin^{2} \zeta }{ q^{2} + m^{2}_{2}} \right]
         LL     \right. \nonumber \\
& & +
         \left[ \frac{sin^{2} \zeta }{ q^{2} + m^{2}_{1}}  + \frac{cos^{2} \zeta }{ q^{2} + m^{2}_{2}} \right]
         RR  \nonumber \\
& & +  \left.  sin \zeta cos \zeta   \left[ -  \frac{1}{ q^{2} + m^{2}_{1}} + \frac{1 }{ q^{2} + m^{2}_{2}}\right]
   \left[ LR + RL \right]
\right\}
\end{eqnarray}

\noindent One notes immediately that for small momentum-transfer q
the left-handed coupling will dominate (as readily observed).  For
high values of the momentum-transfer however ($q^2 \cong m^2_2$)
the interaction becomes parity-symmetric.  Traces of this
"primordial" recovery of mirror-symmetry could eventually be
observed even in the laboratory, at small $q^2$ , as a small
deviation from full parity-violation.

\noindent The first investigations of this eventuality are due to M.A.B. Beg
et al \cite{Beg77} \cite{Holstein77} \cite{Carnoy88}.

\noindent Early decay-asymmetry measurements of $Ne^{19}$ and of the neutron
seemed to indicate a deviation from the Standard Model at the 2.5
$\sigma$-level which could have been explained by the interplay of
a right-handed gauge-boson in the 250 GeV mass-region
\cite{Gaponov90} \cite{Carnoy92}. The conclusion drawn from the
$Ne^{19}$-asymmetry is very sensitive to the ft-value of the
transition ; the most precise result of the neutron-asymmetry
measurements which dominated this conclusion was since modified
(cfr 2.3.1) illustrating the difficulty of absolute
parity-symmetry tests.

\noindent Prompted  by this observation, relative parity-violation
tests were initiated.  The comparison of the longitudinal
positron-polarisation of Fermi- and Gamow-Teller emitters
\cite{Wichers87} \cite{Carnoy91} gave strong constraints in the ($\delta$
- $\zeta$)-plane but did note exclude the possible non-zero value of
$\delta$ .  This is illustrated in Fig. 1 which shows also the
other exclusion regions provided by the most up-to-date values of
both absolute and relative measurements.  This figure shows also
the constraint obtained in the beta polarization/asymmetry
measurements discussed in section 2.3.2.1.

\noindent The exclusion plot combining all nuclear
beta-decay tests is shown (at 90 $\%$ CL) in Fig. 2. No indication
for a predominantly right-handed gauge boson can be noted, the
lower limits of its mass being 320 GeV (at 90 $\%$ CL).  Let us
note the ISOLDE-experiment mentioned in section 2.3.2.1 aims at
improving this limit to about 450 GeV.

\noindent This can be compared to the interesting limits deduced
from the energetics of SN1987A which (provided the right-handed
neutrino is light enough) could not accomodate right-handed bosons
heavier than about 500 GeV (unless they are in the inaccessible
multi-TeV region) \cite{Mohapatra89}.

3.3.1.2. Muon-decay

\noindent For the indications that muon-decay observables provide
on manifest left-right symmetric models we recall the references
given in section 2.1 and the note of J. Govaerts
\cite{Govaerts92}. A mesurement of the positron asymmetry near the
end-point of the spectrum provided a value of $P_{\mu} \xi \delta / \rho$
= 0.99790 +/- 0.00088 \cite{Jodidio67}, 2.4 standard deviations away
from the Standard Model value of 1. Considering however the
possible systematic effects involved in this absolute measurement,
the authors do not consider this deviation as genuine and (for $\xi = 0$
) derive a mass limit of 430 GeV to the right-handed gauge-boson.
The TRIUMF experiment E614, also an absolute one, has a precision
aim of 800 GeV.

\noindent The relative asymmetry-polarization experiment PSI-R
97-06 is also sensitive to right-handed gauge-bosons (cfr.
proposal PSI R-97-06 and the quoted note of J. Govaerts).
Considering that ($\xi''/\xi \xi')- 1 = 4 \delta^2$, a first phase
of the experiment (a relative one !) is expected to provide
already an upper limit of 500 GeV.

3.3.1.3. Constraints of other origin

\noindent The $K_L - K_S$ mass-difference constrains the
right-handed gauge boson mass in manifest left-right symmetric
models to 1.6 TeV \cite{Beall82} or even above 3 TeV
\cite{Bigi86}.

\noindent Direct collider-searches of a second gauge-boson yielded
also upper limits of 650 GeV (at 95 $\%$ CL) by CDF \cite{Abe95}
and 720 MeV (at 95 $\%$ CL) by DO \cite{Abachi96}.

\noindent Let us note that limits derived from the absence of
neutrinoless double beta-decay \cite{Klapdor} do not apply in our
scenario of light right-handed neutrinos (cfr. 3.3.2).

\subsubsection{Generalized left-right symmetric models}

\noindent In generalized left-right symmetric models the
parameter-space ($\delta , \zeta$) is enlarged allowing for
differences of the gauge-couplings, of the
Cabibbo-Kobayashi-Maskawa matrix elements and of the (massif)
neutrino mixing in the left- and right-handed sectors. In addition
to the review papers and other references already quoted, we wish
to call attention to the note of P. Langacker and S. Uma Sankar
\cite{Langacker89} who show not only that the mass-bound of the
right-handed boson deduced from the $K_L - K_S$ mass difference
can be avoided in these generalized models, but also that the
mass-bound deduced from the absence of neutrinoless
double-beta-decay does not apply for light right-handed neutrinos
(the only ones one could hope to observe in low- and
intermediate-energy experiments !).

\noindent As for the limits deduced from colliders, it was noted
by P. Herczeg (private communcication), that they limit the
product of the production cross-section of the W-boson and its
branching-ratio into lepton-pairs and that this provides a
different combination of the parameters than the beta-decay or
muon-decay observables \cite{Govaerts95_1} \cite{Deutsch94}. We
illustrate in Fig. 3 the effect of this difference assuming
similar values for the Cabibbo-Kobayashi-Maskawa matrices of the
two helicity-sectors, but assuming that the gauge-coupling ratio $g_R / g_L$
differs from unity.  As can be seen, for a ratio larger than 2,
the beta-decay experiments become complementary to the
collider-ones.

\noindent A similar conclusion can be drawn from a scenario where
the gauge coupling constants are identical but the
Cabibbo-Kobayashi-Maskawa matrix elements differ.  In this
scenario, illustrated in Fig. 4, the beta-decay experiments could
exclude right-handed gauge-bosons of a mass similar to (or smaller
than) the known one of 80 GeV. The consistency of such a scenario
with other constrains was not yet explored \cite{Langacker}.

\noindent Nuclear beta-decay tests have to assume light enough
right-handed electron neutrinos.  As the muon-decay tests have to
assume this also for muon-neutrinos, the two types of experiments
are complementary. It should also be noted that in muon-decay the
asymmetry-measurement (PSI R-97.076) test different combinations
of the right-handed neutrino sector \cite{Govaerts95_1} making these
two experiments complementary in generalised left-right symmetric
models.

\section*{Acknowledgments}

\noindent I am grateful to J. Govaerts and P. Herczeg for a critical reading
of the manuscript and to my collaborators J. Govaerts, R. Prieels,
N. Severijns and O. Naviliat-Cuncic for many useful discussions.
E. Thomas was of great help in the analysis and producing the
figures.
I profited also from the comments of P. Langacker, R. Oakes and H.
Klapdor-Kleingrothaus. Many colleagues kindly informed me of their
plans and provided useful details on them.  I am particularly
thankful in this respect to H. Abele, E. Adelberger, L.N.
Bondarenko, T. Bowles, J. Byrne, S. Dewey, J. Doyle, D. Dubbers,
W. Fetscher, A. Garcia, A. Hime, I. Kuznetsov, S. Lamoreaux, D.
Mischke, Cl. Petitjean, D. Pocanic and D. Wright.

\begin{figure}[h]
        \begin{center}
                \psfig{figure=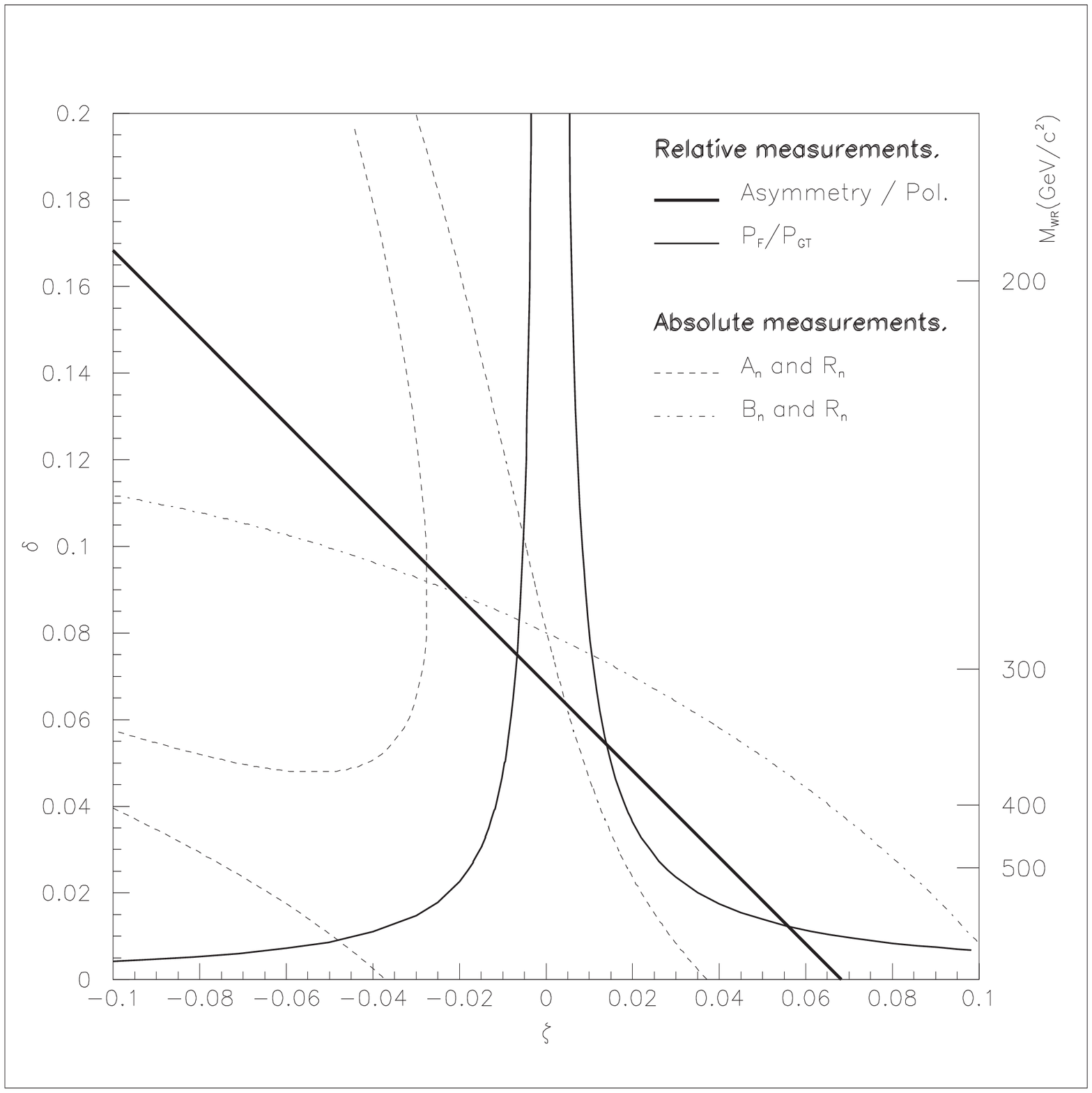,height=15cm}
        \end{center}
        \caption{
        \label{Figure:exclusion} Comparative exclusion plots of the manifest left-right
        symmetric model parameters provided by various absolute and relative nuclear
        beta-decay tests at their present levels of precision (90\% CL). $R_{n}$ is the
        ratio of the neutron-and Fermi decay ft-values}
\end{figure}

\begin{figure}[h]
        \begin{center}
                \psfig{figure=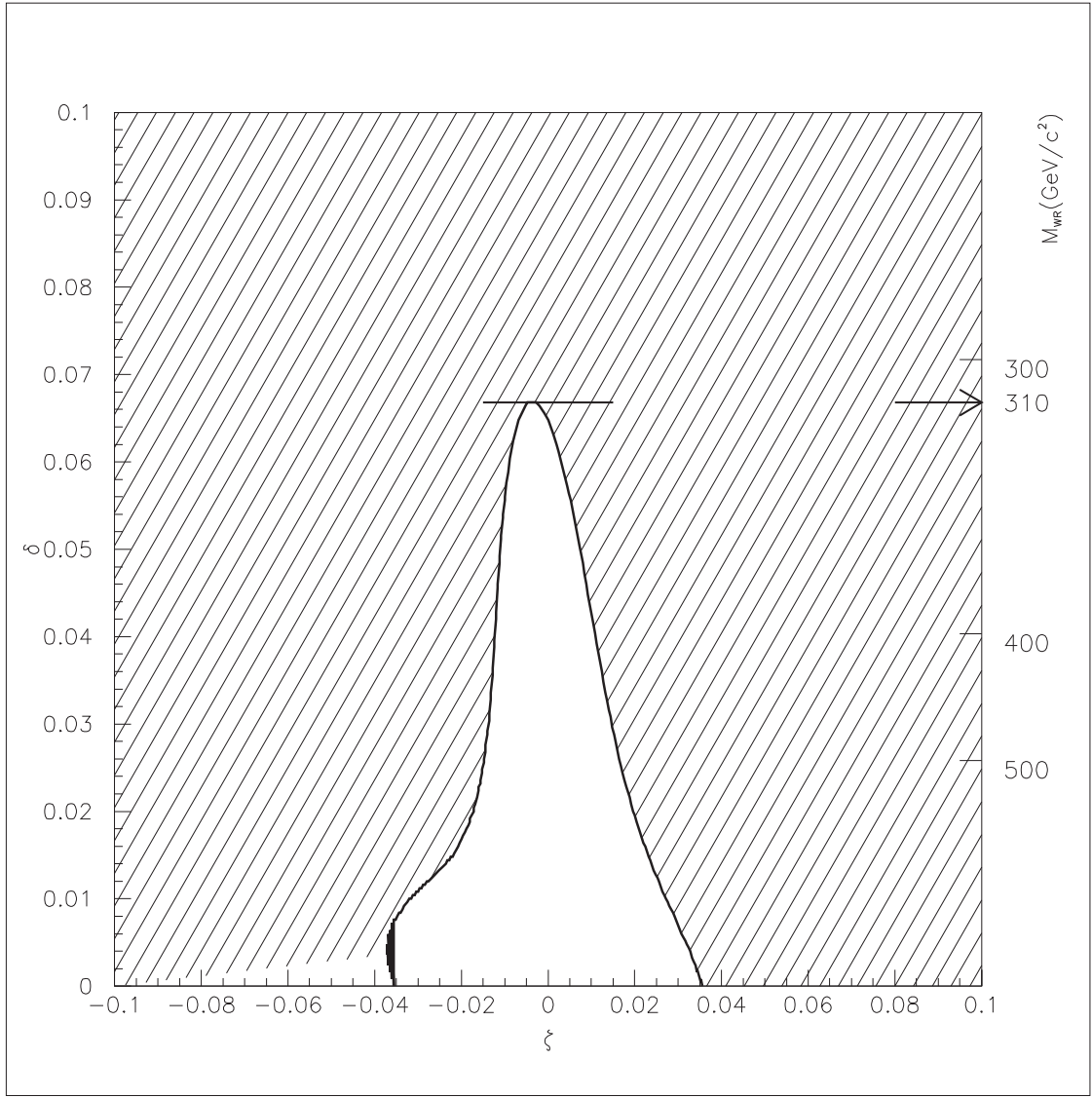,height=15cm}
        \end{center}
        \caption{
        \label{Figure:combination} Combination of the constraints presented in Figure \ref{Figure:exclusion}
        on the 90 \% CL. The lower limit of the right handed gauge boson mass (independently of the
        mixing parameter) is represented by an arrow.}
\end{figure}

\begin{figure}[h]
        \begin{center}
                \psfig{figure=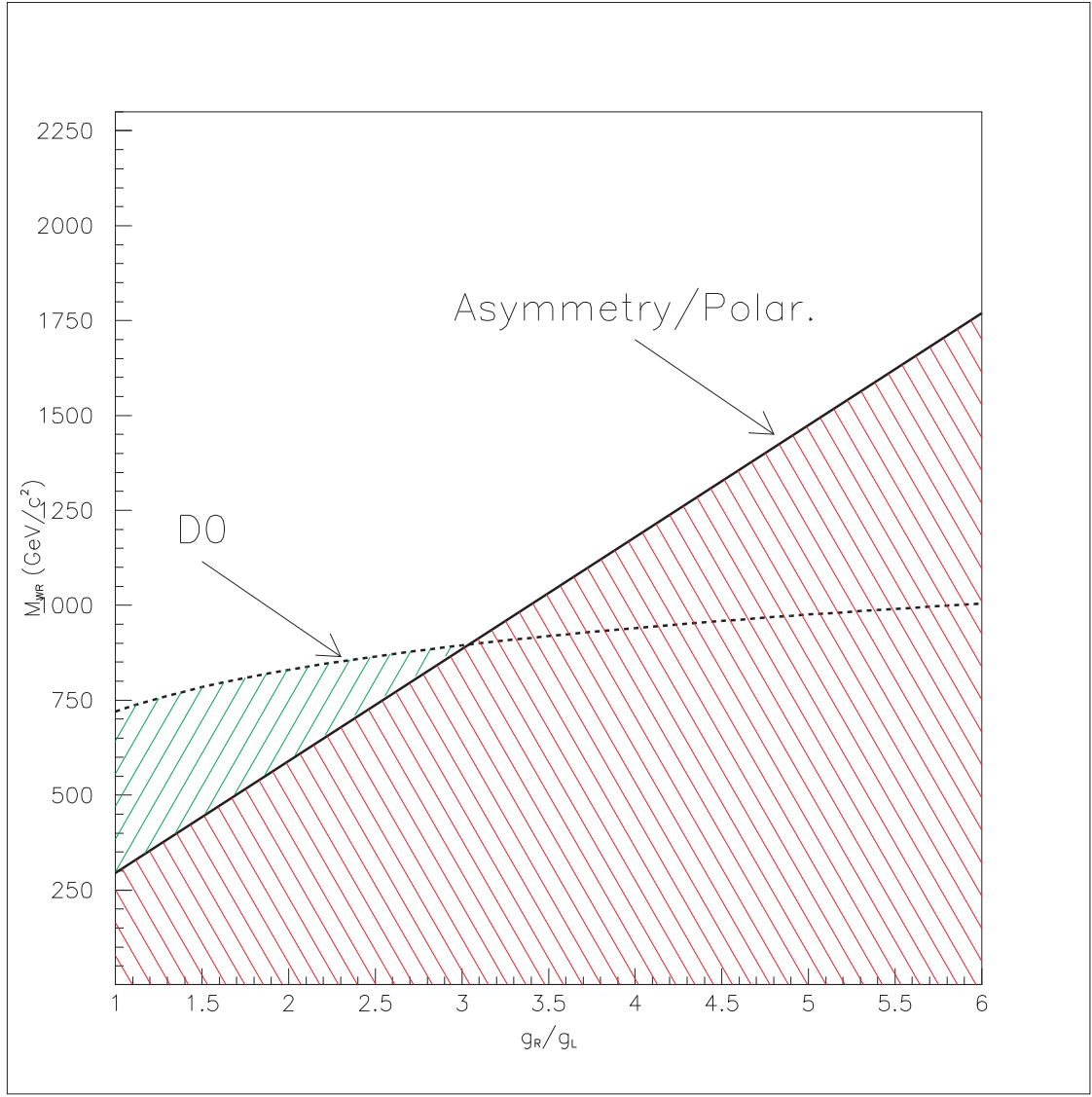,height=15cm}
        \end{center}
        \caption{
        \label{Figure:asymmetry} Right-handed gauge-boson limits (at 90\% CL) deduced from
        collider-experiments and nuclear beta-decay of the type discussed in 2.3.2.1
        (for $\zeta = 0$) as a function of the gauge-coupling ratio of the right (-left)-handed
        sectors.}
\end{figure}

\begin{figure}[h]
        \begin{center}
                \psfig{figure=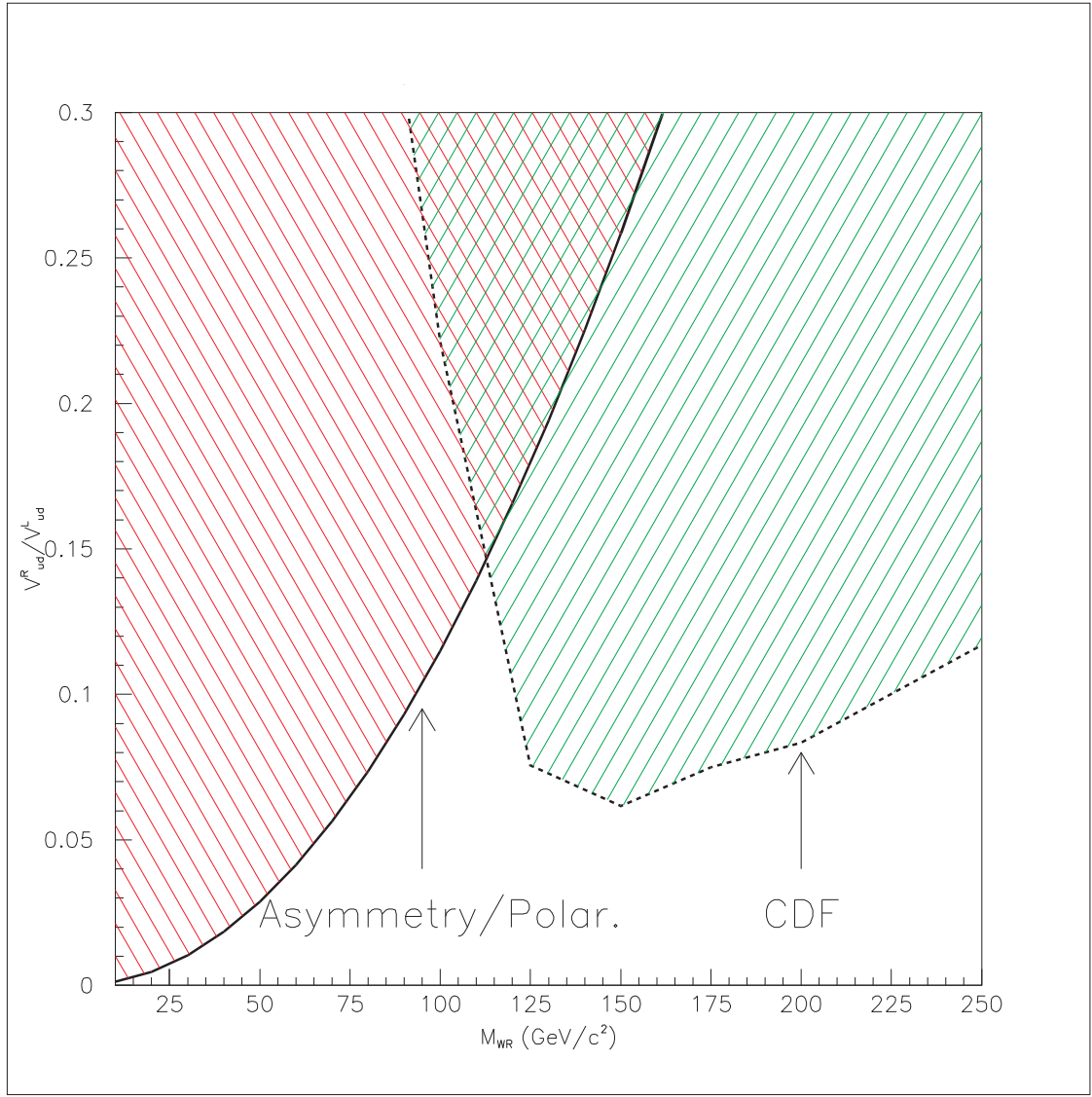,height=15cm}
        \end{center}
        \caption{
        \label{Figure:cabibbo} Exclusion-plots of the right(left)-handed Cabibbo-Kobayashi-Maskawa
        matrix-element ratio as a function of a (light) right-handed gauge-boson mass.}
\end{figure}

\end{document}